\documentclass[a4paper,fleqn,usenatbib]{mnras}
\usepackage{graphicx}
\usepackage{amsmath}
\usepackage{amssymb}
\usepackage{upgreek}
\usepackage{longtable}
\usepackage{hyperref}

\title[Polarimetry of 600 radio pulsars]
{Polarimetry of 600 pulsars from observations at 1.4~GHz
with the Parkes radio telescope}

\author[Johnston \& Kerr]
{Simon Johnston$^{1}$\thanks{email: Simon.Johnston@csiro.au} and
{Matthew Kerr$^{1,2}$}
\\
$^{1}$CSIRO Astronomy and Space Science, Australia Telescope National Facility, PO Box 76, Epping, NSW 1710, Australia\\
$^{2}$Space Science Division, Naval Research Laboratory, Washington, DC
20375-5352, USA\\
}
\date{Accepted \today. Received \today; in original form \today}

\pubyear{2017}

\begin{document}
\label{firstpage}
\pagerange{\pageref{firstpage}--\pageref{lastpage}} 
\maketitle

\begin{abstract}
Over the past 13 years, the Parkes radio telescope has observed a large number
of pulsars using digital filterbank backends with high time and frequency 
resolution and the capability for Stokes recording. Here we use archival
data to present polarimetry data at an observing frequency
of 1.4~GHz for 600 pulsars with spin-periods ranging from 0.036 to 8.5~s.
We comment briefly on some of the statistical implications from the data and
highlight the differences between pulsars with high and low spin-down 
energy. The dataset, images and table of properties for all 600 pulsars
are made available in a public data archive maintained by the CSIRO.
\end{abstract}


\section{Introduction}
The first pulsar discovery was made nearly 50 years ago \citep{hbp+68}
and it was quickly realised that pulsars were often highly
polarized \citep{ls68}. One of the defining characteristics of pulsars
is the long-term pulse profile stability combined with the complexity of the 
profiles in both total intensity and polarization. The polarimetry
of pulsars leads to an understanding of the geometry of the star, an 
estimate of the height of the radio emission, clues as to the emission physics
and the overall structure of the pulsar beam.

It became apparent that the radio emission
is strongly tied to the magnetic field lines and the position angle
of the linear polarization is determined by the angle of the magnetic
field line as it sweeps past the line of sight \citep{kom70}. This in turn leads
to an understanding of the geometry of the star, in particular the angle
between the rotation and magnetic axes ($\alpha$) can be estimated 
via the `rotating vector model' \citep{rc69}. Furthermore, relativistic
effects can offset the position angle swing from the total intensity
profile and this can be used to measure the height of the radio
emission \citep{bcw91}. The total intensity profiles themselves are used to 
infer the structure of the emission beam, with \cite{ran90,ran93} favouring
a structured model with `core' and `cone' emission whereas \cite{lm88}
prefer a more patchy, random emission beam. Traditionally it was thought
that lower frequencies arise from higher in the magnetosphere than higher
frequencies (see e.g. \citealt{cor78}) but more recent evidence suggests
that emission arises over a wide range of heights in a given pulsar
at a single frequency \citep{gg03,kj07}.

The position angle swing as a function of pulse phase is often interrupted by 
jumps of 90\degr\ and this led to the idea that emission was present in two 
orthogonal modes \citep{em69,brc76}.
Whether the modes are disjoint or superimposed has
been debated by observers \citep{ms98,es04} and 
theorists alike \citep{gss91,mmkl06,vt17} and the question remains unresolved.

Finally the origin of the polarization remains unclear with circular
polarization discussed empirically and theoretically by e.g. \cite{rr90}
and \cite{ml04}. Refraction and propagation effects in the magnetosphere
and conversion between the modes in a `polarization limiting region'
has been put forward by e.g.  \cite{pl00} and \cite{pet06} as an
explanation for features seen in linear polarization.
Cyclotron absorption in the pulsar magnetosphere \citep{bs76} may play
a role in both the profile shape and the polarization characteristics.
\citet{lm01} show that the trailing part of profiles are suppressed 
relative to the leading part and discuss the possibility that one of the
polarized modes may also be absorbed, preventing mode mixing and leading to a 
high degree of polarization in the profile as a whole.

Publication of polarimetry for a large sample of pulsars is therefore
vital for tackling the problems outlined above. Observations made
at 1.4~GHz have been published for
300 pulsars with the Lovell telescope in the UK \citep{gl98},
98 pulsars with the Arecibo telescope in the USA \citep{wcl+99} and
66 pulsars with the Parkes telescope in Australia \citep{mhq98}.
Recent polarimetric observations at lower frequencies include those
compiled by \citet{hr10} and \citet{mbm+16}.
Here, we provide data on 600 mostly southern hemisphere pulsars observed
with the Parkes radio telescope over the past decade.
The aim of this paper is threefold. First to make available the entire dataset
of southern hemisphere pulsars, flux and polarimetrically calibrated,
in {\sc psrfits} format \citep{hvm04} for use by other
researchers. Secondly, to provide thumb-nail images for all the pulsars
and finally to provide in tabular form the polarization characteristics
for this large sample of pulsars.
\begin{figure}
\includegraphics[width=8cm]{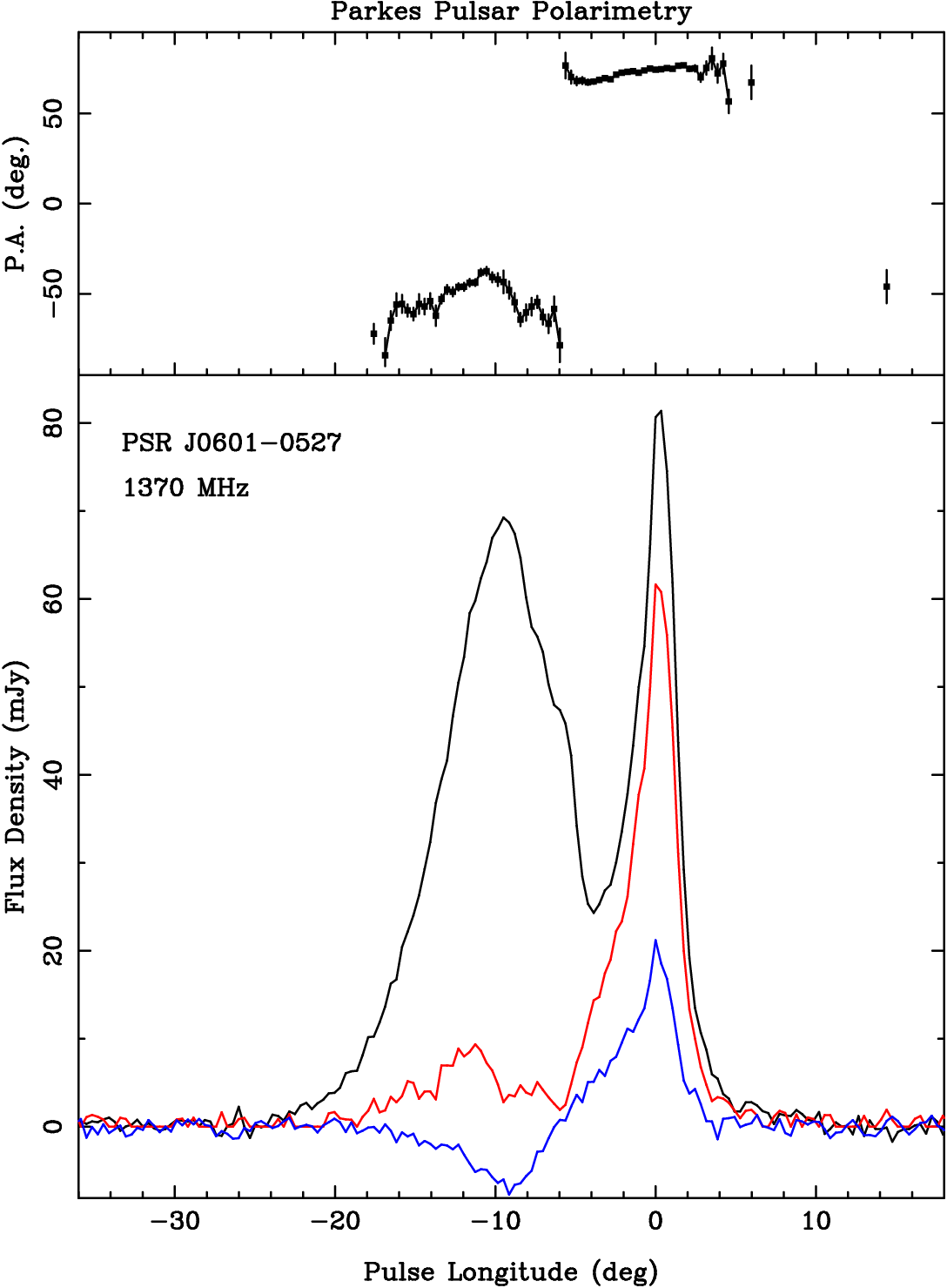}
\caption{PSR~J0601--0527 in full polarization. In the lower panel the
black line denotes Stokes I, the red trace shows the linear polarization
and the blue trace the circular polarization. Left-hand circular polarization
is defined to be positive. The top panel shows the
position angle of the linear polarization, corrected to infinite frequency
using the RM listed in the table. Position angles are only plotted when
the linear polarization exceeds 3 sigma. The zero point of pulse
longitude is set to the peak of the total intensity profile.}
\label{J0601}
\end{figure}

In Section~2 we outline the observations, section~3 deals with the data
reduction, section~4 presents the results and section ~5 briefly discusses
the implications in general terms. Section~6 informs readers of how to
access the data.

\section{Observations}
\begin{table}
\caption{Origin of the archival pulsar observations}
\label{obstable}
\begin{tabular}{lllcrc}
ID & PI & Obs Dates & Rx & Pulsars & Ref\\
\hline & \vspace{-3mm} \\
P574 & Johnston & 2008-2017 & MB & 301 & 1 \\
P535 & Johnston & 2006-2007 & H-OH & 149 \\
P236 & Han & 2006-2007 & MB & 54 & 2\\
P630 & Johnston & 2009-2012 & MB & 48 & 3\\
P439 & Johnston & 2004-2005 & H-OH & 21 & 4\\
P875 & Jankowski & 2014 & MB & 21 & 5\\
Misc & & & & 6 \\
\hline
\end{tabular}

1. \citet{jw06,wj08b,wjm+10,aaa+13,rwj15a}.
2. Profiles unpublished, data presented in \citet{hmvd17}.
3. \citet{tjb+13}.
4. \citet{jhv+05,jkk+07}.
5. Profiles unpublished, data presented in \citet{jan17}.
\end{table}
The Parkes radio telescope is a 64-m dish located near Parkes, NSW, Australia.
Functional since 1963, the telescope has had a long history of observations of
radio pulsars since their discovery in 1967.
In 2004, the telescope acquired digital backend systems to replace the
aging analogue spectrometers. These Digital Filterbanks (DFBs) were
FPGA-based correlators, capable of recording full Stokes information over
a large number of frequency channels. During pulsar observations, the 
data are folded modulo the pulsar's topocentric spin frequency to
form a `sub-integration' of typical duration 10-30~s. These sub-integrations
are combined until the desired total observing length had been reached.

At observing frequencies near 1.4~GHz two main receivers have been available
over the past decade, the multibeam receiver \citep{swb+96} and the H-OH 
receiver \citep{tgj90}. For the purposes described here,
both receivers are generally used 
over 256~MHz of bandwidth from 1240 to 1496~MHz.  Typically the bandwidth
is divided into 512 or 1024 frequency channels each of width 0.25 or 0.5~MHz.
For both systems, it is possible to inject a square-wave
signal of known periodicity directly into the feed for calibration purposes.
For further details of the receiver and backend systems see \citet{mhb+13}.

In this paper we concentrate on `normal' pulsars with spin periods
greater than 30~ms and which are not considered to have been recycled.
For a systematic view of the polarimetry of millisecond pulsars
in the southern hemisphere see \cite{dgm+15}.
The most recent publication of a systematic survey of southern pulsars 
in polarization dates back nearly two decades \citep{mhq98} and
have poorer time and frequency resolution than the data collected here.

The observations span a time period from 2004 November to 2017 February
and data were taken from the public Data Archive 
Portal (\citealt{hmm+11}; \url{http://data.csiro.au}).
A total of 600 pulsars were used for this work, many of them 
observed multiple times.
Table~\ref{obstable} shows the origin of the data, listing
the project identifier (ID) and principal investigator (PI), the dates of the 
observations, the receiver used (Rx; MB denotes the multi-beam receiver),
the number of pulsars and a reference to previously published material.

\section{Data Reduction}
\begin{figure}
\includegraphics[width=8cm]{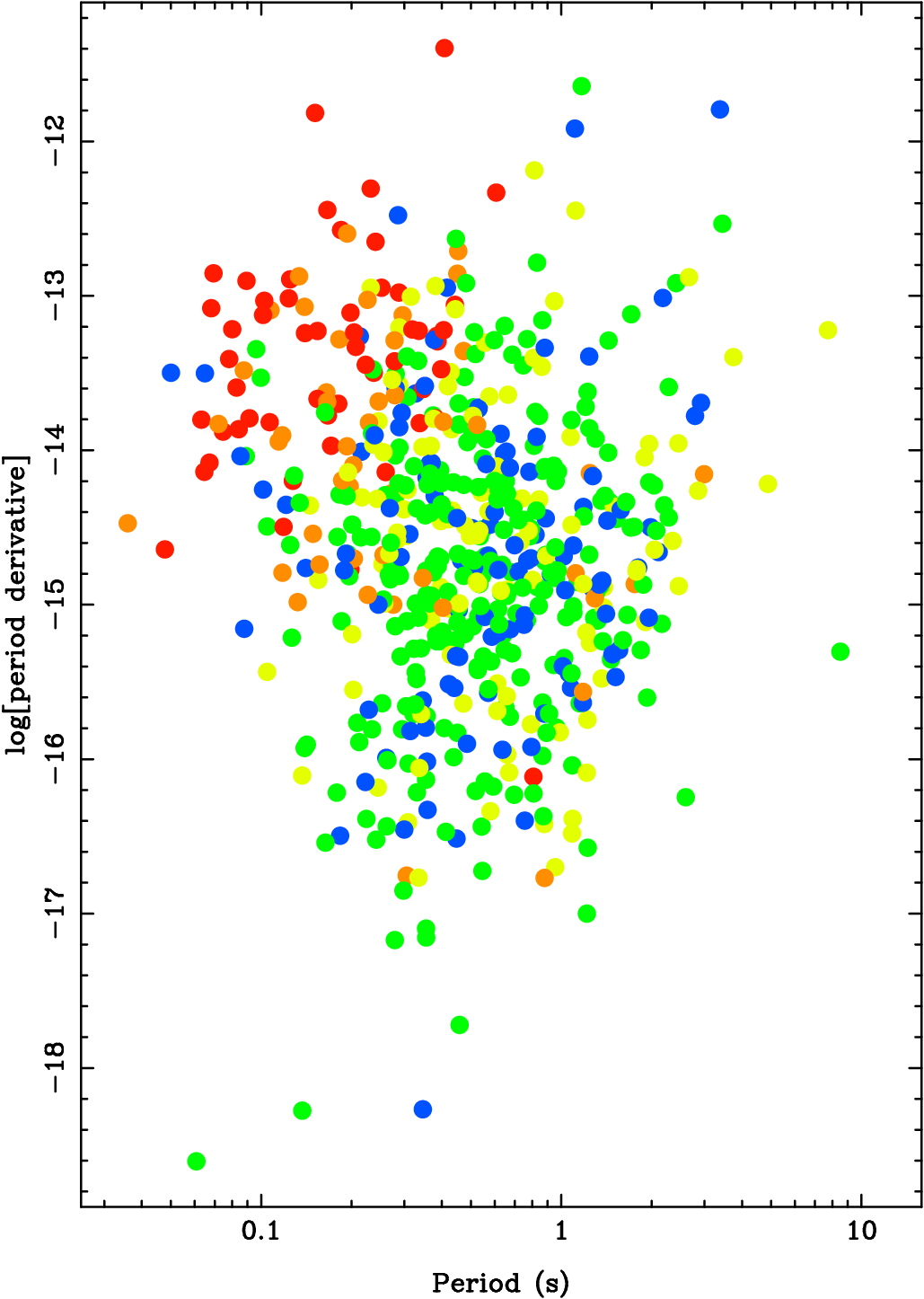}
\caption{The period-period derivative plane for 594 pulsars. The colour
represents the fraction of linear polarization with red the highest
and blue the lowest.}
\label{ppdot}
\end{figure}
The entire dataset under consideration here has been reprocessed for
this paper. We outline the basic steps below.
\subsection{Phase and gain calibration}
A calibrator observation consists of pointing the telescope at a patch
of blank sky and injecting a square-wave with a 50\% duty cycle and known
periodicity directly into the feed. The equivalent flux density of
the calibrator is $\sim$1~Jy.
A calibrator observation is first located as close in time and in
sky location as possible to the pulsar observation. For the vast majority
of observations, the calibrator was observed within one hour of the pulsar
observation and in the same area of sky. Frequency channels which are corrupted
(either edge channels or channels with interference) are identified and 
deleted from the data. Then, the relative gain between the two linear
polarization probes are determined along with the phase offset between them
on a channel-by-channel basis.

\subsection{Flux density calibration}
Flux density calibration is carried out on a regular basis using the source
Hydra~A as a reference. Hydra~A has a flux density of 43.1~Jy at 1.4~GHz.
For the majority of the observations described here,
flux density calibration measurements were made within two weeks of the
pulsar observations. Results from the flux density calibration are applied 
to the pulsar data on a per-channel basis
to convert from digitiser counts into units of Jy.

\subsection{Receiver characterisation}
The multibeam receiver shows a modest level of cross-polarization
and/or non-orthogonality. We characterize this following the method of 
\citet{vans04} where long observations, over a range of parallactic
angles, of a highly polarized source (typically PSR~J0437--4715) are used
to model the receiver response. Such observations are carried out
every few months.  We compute and apply the resulting
calibration solutions with the {\sc psrchive} routine {\sc pcm}.
For the H-OH receiver, each pulsar was observed at least twice with
the feed rotated by 90\degr\ between observations to minimise the impact of 
the already small impurities in the receiver (see details in
\citealt{jhv+05}).

\subsection{Pulsar data reduction}
The pulsar data are first corrected for gain and phase by applying the 
solutions obtained from the calibrator and then converted to Jy by applying 
the flux calibration solution. Note that any corrupt frequency channels
identified in the previous step are also then flagged in the pulsar data.
Receiver calibration solutions are then applied.
Next, the pulsar observation is examined, and corrupt sub-integrations are 
flagged and removed from the data.
Finally, the pulsar data is averaged in both time and frequency, taking
into account the rotation measure (RM) and dispersion measure (DM)
of the pulsar.  The majority of the RM values were obtained from the large 
samples reported in \citet{hml+06}, \citet{njkk08} and \citet{hmvd17}. 
Several observations of the same pulsar may be available. If so, the
(flux and polarimetrically calibrated) observations are added together after 
alignment via cross-correlation.
The final output contains a Stokes I,Q,U,V datastream
covering the pulsar period with a resolution of 1024 or 512 time samples.

\subsection{Polarization conventions}
The polarization data have absolute calibration using
the conventions outlined in \cite{vmjr10}. Position angles are defined
as increasing counter-clockwise on the sky (see \citealt{ew01}).
Circular polarization in pulsar astronomy uses the IEEE convention
for left-hand and right-hand, a convention which differs from the one
adopted by the IAU and which most radio interferometers apply.
In the data presented here, left-hand circular
polarization is positive as per the IEEE and pulsar convention.
A sample pulsar is shown in Figure~\ref{J0601}.

\begin{figure*}
\begin{center}
\begin{tabular}{cc}
\includegraphics[width=8cm,angle=0]{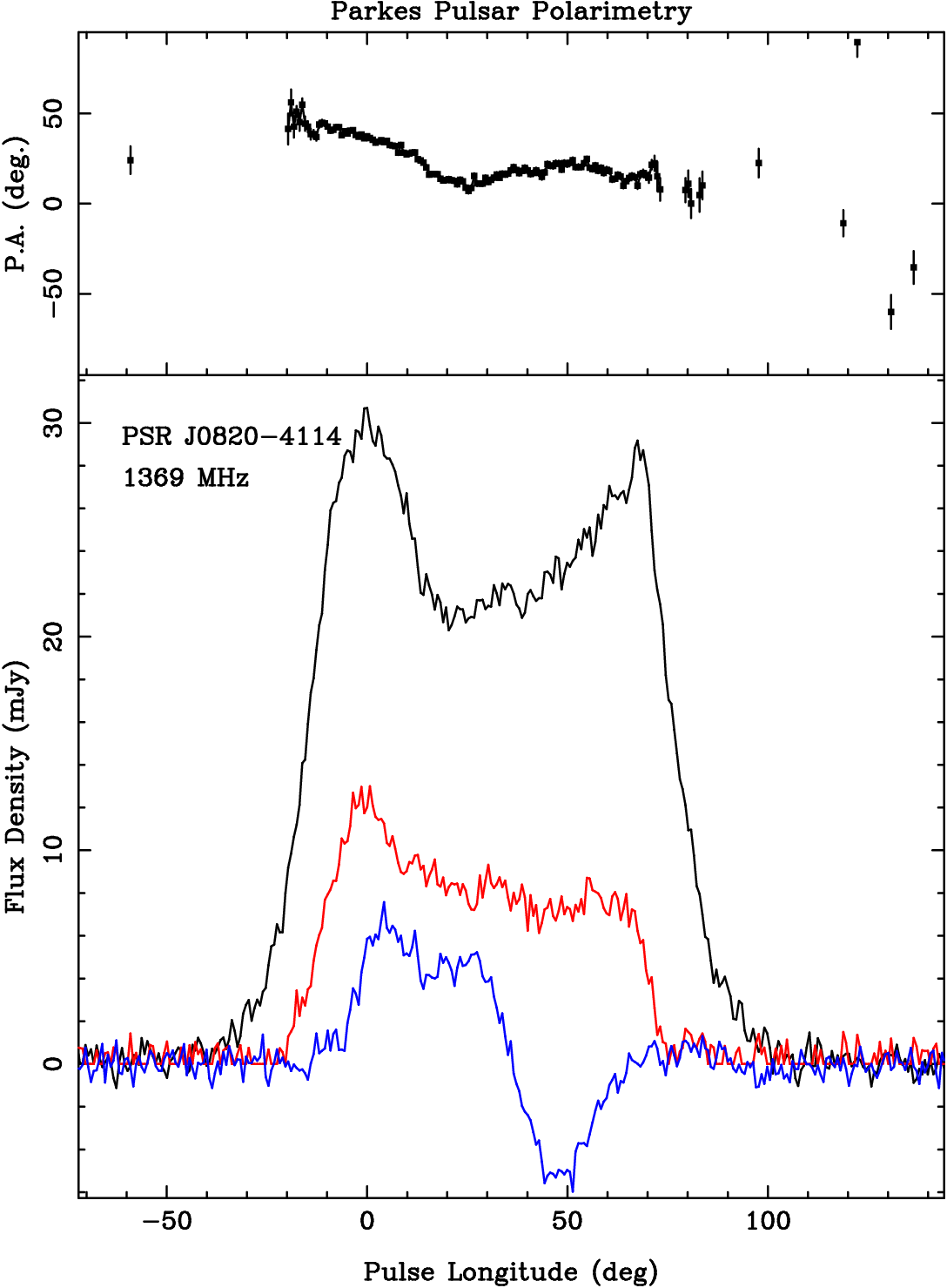} &
\includegraphics[width=8cm,angle=0]{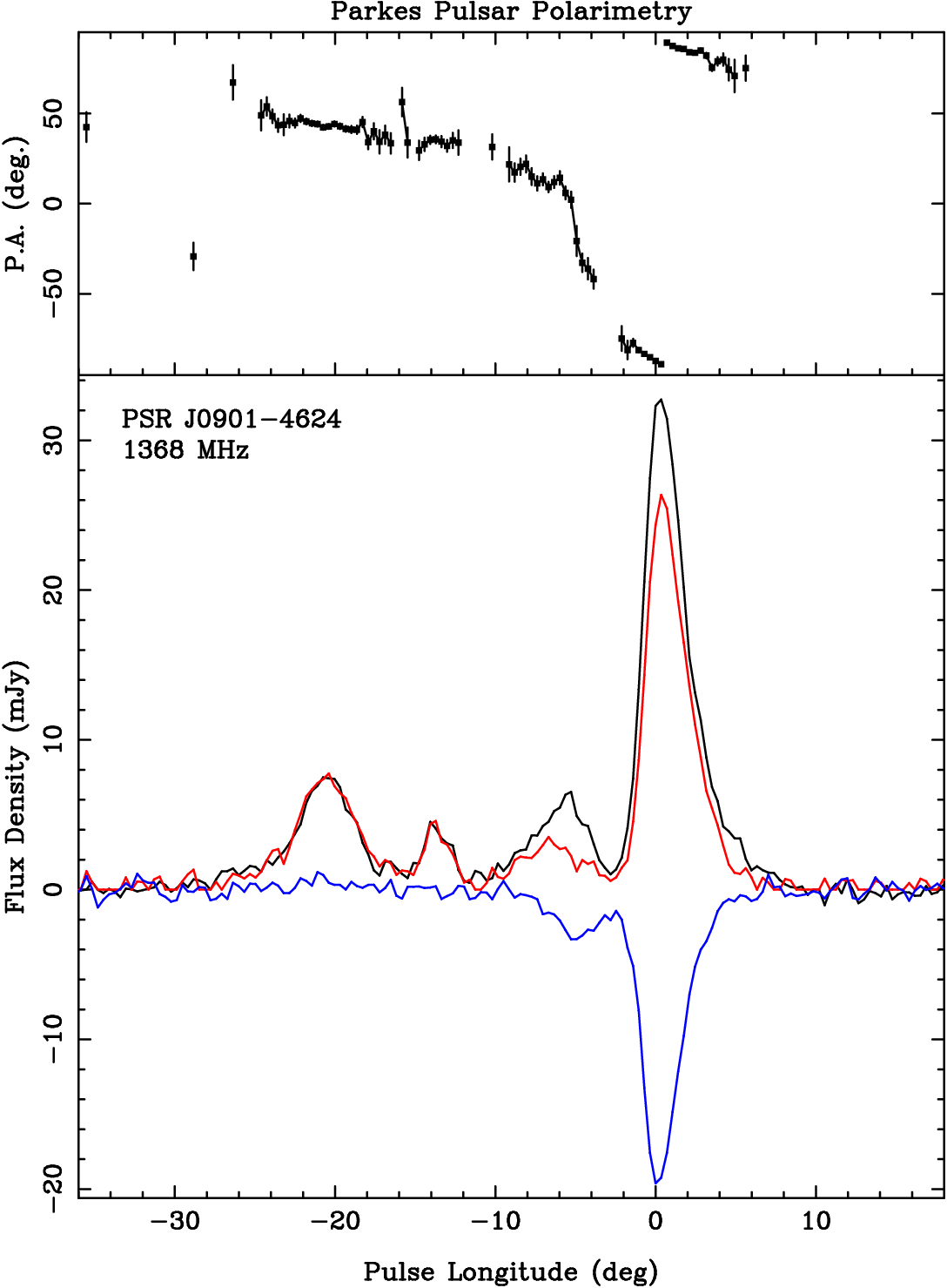} \\
\includegraphics[width=8cm,angle=0]{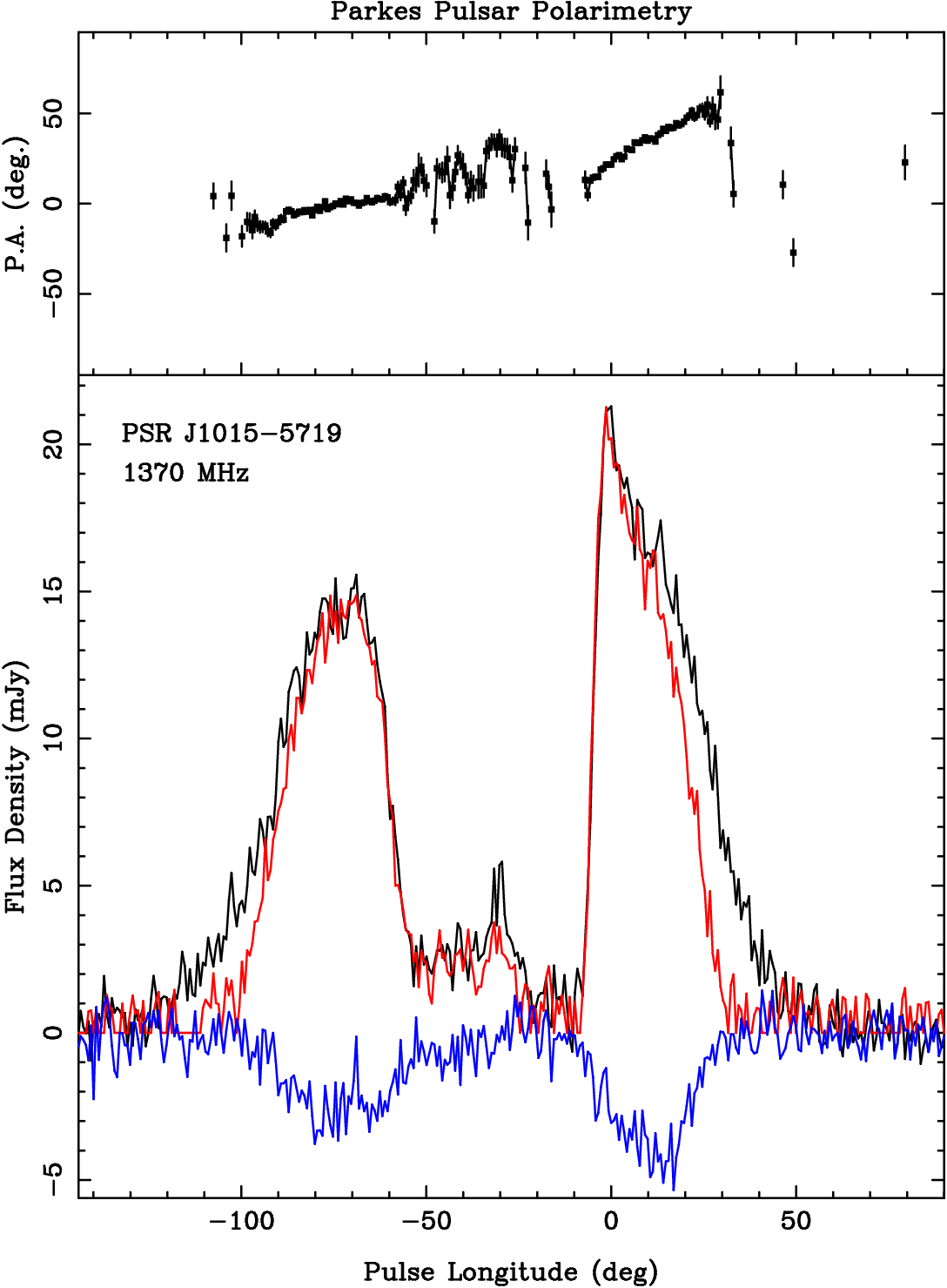} &
\includegraphics[width=8cm,angle=0]{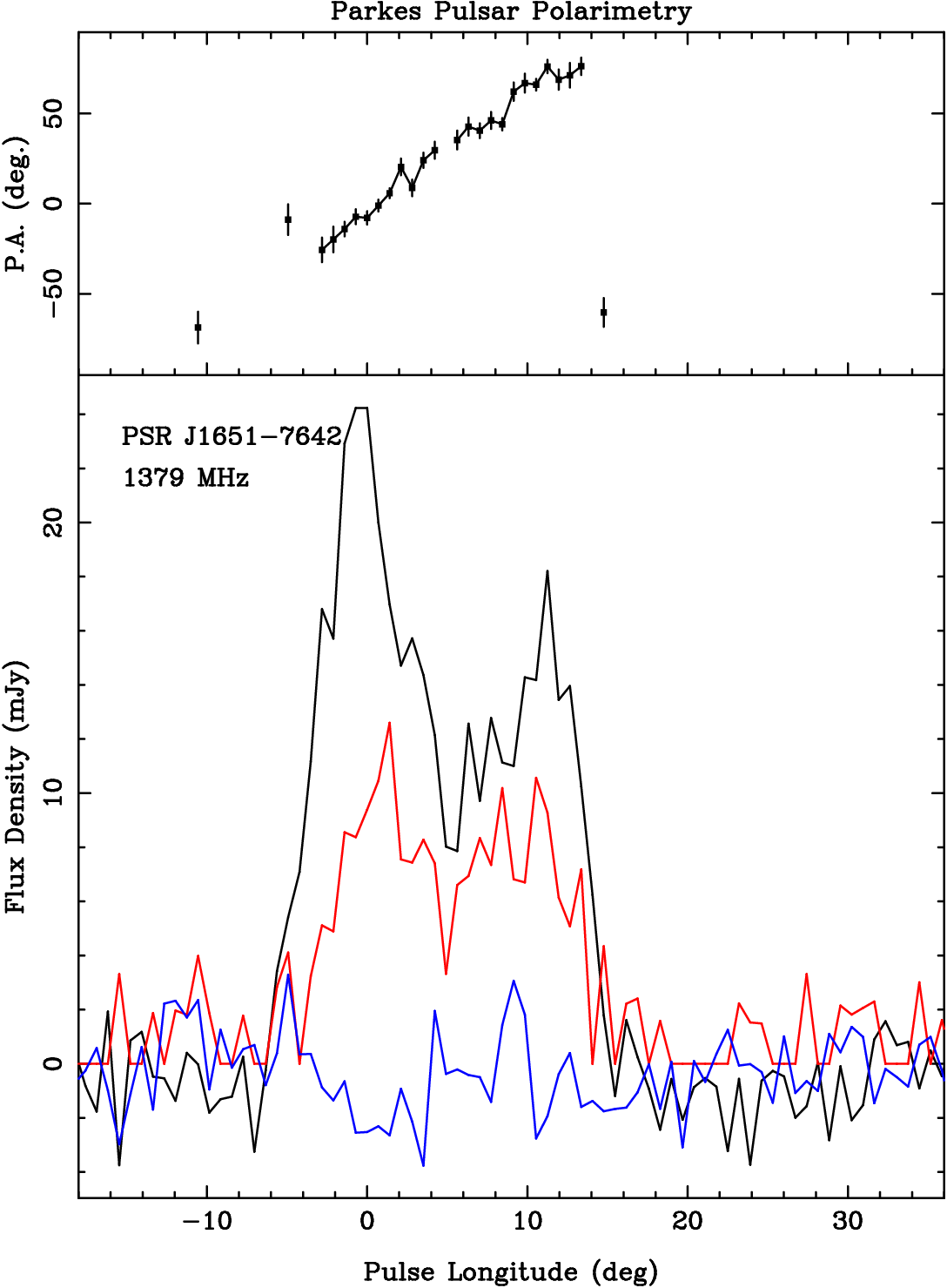} \\
\end{tabular}
\end{center}
\caption{Example pulsars from the discussion.
Top left is PSR~J0820--4114 an example
of a low $\dot{E}$ pulsar with a large width, top right is PSR~J0901--4624 which
has a large fractional circular polarization.
Bottom left is an example of a wide-double, high $\dot{E}$ pulsar
PSR~J1015--5719 and bottom right shows a rare example of a 
pulsar with high linear fraction in spite of low $\dot{E}$.
See also the caption to Figure~\ref{J0601}.}
\label{4psrs}
\end{figure*}
\section{Results}
The distribution of 594 pulsars on the period-period derivative
($P-\dot{P}$) diagram is shown in Figure~\ref{ppdot} (6 pulsars have no
measured $\dot{P}$).  The periods range from 35~ms
to 8.5~s and there is 7 orders of magnitude coverage in $\dot{P}$.

Table~\ref{bigtable} lists the polarization properties of the 
first 25 pulsars (the full table is available through the on-line material).
The table shows the pulsar parameters, spin period ($P$), dispersion
measure (DM) and the spin-down energy ($\dot{E}$) given by
\begin{equation}
\dot{E} = 4\pi^2 I \,\,\,\dot{P}\,\,\,P^{-3}
\end{equation}
where $I$ is the moment of inertia of the star (taken to be
$10^{45}$~g\,cm$^{2}$). The rotation measure
(RM) as used in the data reduction is listed.
The table then lists then number of
bins (nbin) across the pulsar profile and the off-pulse (i.e. the noise)
root-mean-square ($\sigma_I$) in mJy per bin.
The flux density at 1.4~GHz ($S_{1.4}$) is listed along with the width
of the profile (in degrees) at 50\% of the peak ($W_{50}$).
Taking into account systematic effects in the calibration, we estimate the 
error on $S_{1.4}$ to be 10\%. The error on $W_{50}$ is $\lesssim$0.5\degr\
when the signal to noise ratio is high, but quickly becomes large once
the signal to noise ratio drops below 10.
Finally, the percentage of linear polarization ($L/I$),
of circular polarization ($V/I$) , and of the absolute value of the circular 
polarization ($|V|/I$) is given along with the error bar.
For computation of the fractional linear polarization we follow \cite{ew01},
in particular their Equation~11 is used to debias $L$.
In order to compute the $|V|/I$ we have 
followed the prescription given in \cite{kj04} which ensures that 
the off-pulse baseline retains a mean of zero.
The error bar is given by the quadrature sum of the formal fits to the profile,
taking into account $\sigma_I$, and our estimate of the systematic errors (3\%).
\begin{figure}
\includegraphics[width=6cm,angle=270]{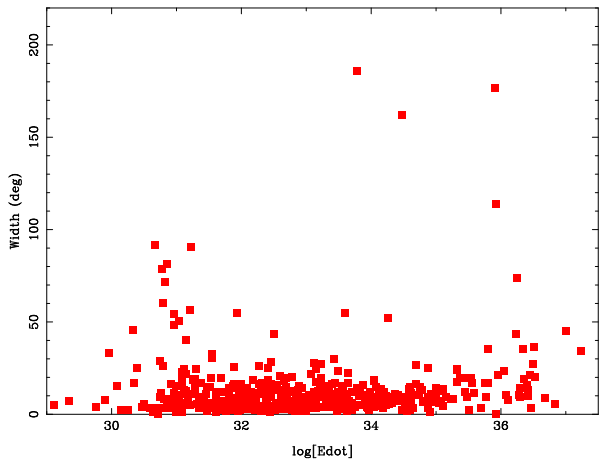}
\caption{Spin-down energy ($\dot{E}$) versus pulse width ($W_{50}$) for 501 pulsars.}
\label{edot-w50}
\end{figure}
\begin{figure}
\includegraphics[width=6cm,angle=270]{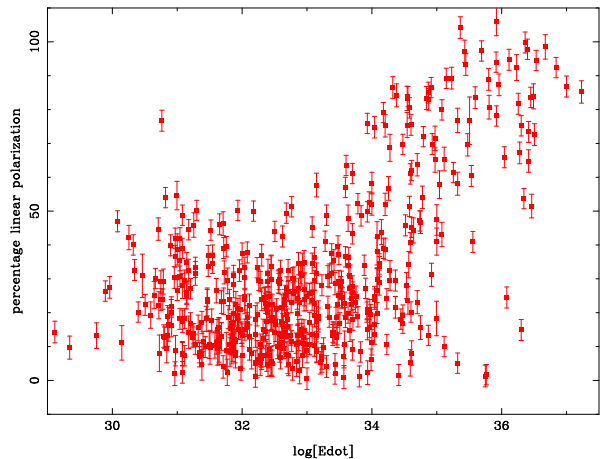}
\caption{Spin-down energy ($\dot{E}$) versus percentage linear polarization for 501 pulsars.}
\label{edot-linear}
\end{figure}

The last column contains a flag. ``W" implies the pulsar has a peak flux
density less than 15 times the rms; $W_{50}$ and the fractional 
polarization values should be treated with caution. There are 69 such pulsars
in the sample.
``R" denotes a pulsar for which we could obtain no RM either because the
pulsar is weak or the fractional linear polarization is intrinsically low.
The RM is set to zero for these cases. A total of 55 pulsars fall into 
this category (of which 25 are also classified as ``W").
``S" denotes a pulsar showing strong (clearly visible by eye) 
scatter-broadening of the profile
due to the interstellar medium. In this case, the true pulse width will
be smaller than the measured width and the measured fractional polarization will
be lower than the true value. There are 32 pulsars with this flag.
``N" denotes pulsars that are not flux calibrated. There are 14 pulsars in this
category; their flux densities have been set to zero in the table.
``I" denotes the 20 pulsars with an interpulse.

\section{Discussion}
In the discussion that follows, pulsars with a ``W" or ``S" flag 
have been excluded from the analysis and the figures.
Sample pulsars which form part of the discussion can be found 
in Figure~\ref{4psrs}.

Figure~\ref{edot-w50} shows $W_{50}$ against $\dot{E}$. There are four
groupings which stand out in the figure. First, the bulk of the
pulsar population have $W_{50} \lesssim 25\degr$; a slight rise in  $W_{50}$
to higher values of $\dot{E}$ can be seen. Secondly there are a group
of pulsars with $\dot{E}<10^{32}$~erg\,s$^{-1}$ and $W_{50} \gtrsim 50\degr$.
These pulsars have small 
values of $\alpha$, examples include PSRs~J0820--4114 (see Figure~\ref{4psrs}),
J1034--3224 and J1133--625. Thirdly there
are the moderate $\dot{E}$ pulsars with very large $W_{50}$. These
are the interpulse pulsars such as PSR~J0908--4913 \citep{kj08,kjwk10}
where $\alpha$ is close to 90\degr\ and both poles are seen.
Finally, some high $\dot{E}$ pulsars have large widths. These are
the ``wide-doubles'' identified by \cite{jw06} examples of which include
PSRs~J1015--5917 (see Figure~\ref{4psrs}) and J1302--6350. The geometry
of these pulsars and the implications thereof have been examined in
detail by \citet{rwj15b}.

Figure~\ref{edot-linear} shows $\dot{E}$ versus linear polarization fraction
and contains the striking result obtained by \cite{wj08b}.
Below $\dot{E}$ of 10$^{33}$~erg\,s$^{-1}$, only 3 objects have a linear
fraction in excess of 55\%, PSRs~J0108--1431, J1651--7642 and J1805--1504
(see Figure~\ref{4psrs}).
In contrast, above $\dot{E}$ of 10$^{35.5}$~erg\,s$^{-1}$ only 5 objects
have $L/I < 50\%$, PSRs~J1055--6028, J1413--6141, J1646--4346,
J1833--0827 and J1837--0653.
Figure~\ref{ppdot} gives another representation of this effect. Pulsars
at the top left of the diagram are highly polarized (red colour) as
opposed to the remainder of the pulsars which have lower polarization
(green/blue colours).

\begin{figure}
\includegraphics[width=6cm,angle=270]{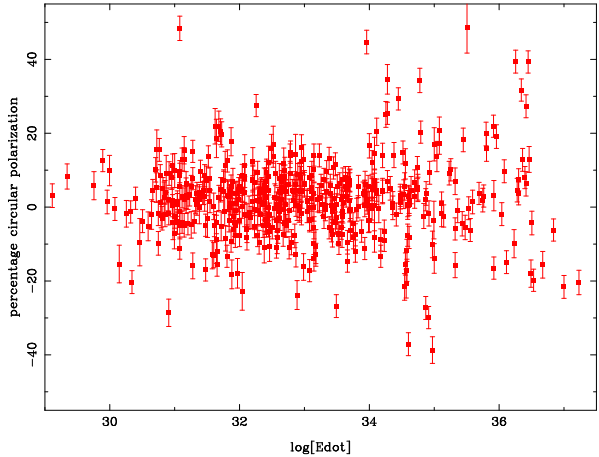}
\caption{Spin-down energy ($\dot{E}$) versus percentage circular polarization for 501 pulsars.}
\label{edot-circ}
\end{figure}

Figure~\ref{edot-circ} also shows some differences in the
fractional circular polarization between the high $\dot{E}$ and the
low $\dot{E}$ pulsars, with the high $\dot{E}$ pulsars having a larger
spread in circular polarization and a higher average.
There is no preference for right or left
handed circular. Only very few pulsars have $V/I \gtrsim 40\%$, examples
are PSRs~J0901--4624 (see Figure~\ref{4psrs}), J1410--7404 and J1915+1009,
and there are no convincing examples of pulsars with $V > L$.

\begin{figure}
\includegraphics[width=6cm,angle=270]{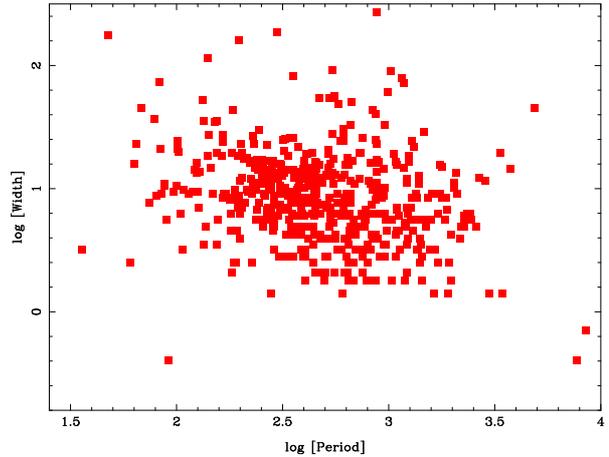}
\caption{Log of the spin period ($P$) versus log of the pulse width 
($W_{50}$) for 501 pulsars.}
\label{pw}
\end{figure}

Figure~\ref{pw} shows the relationship between $P$ and $W_{50}$.
Although there
is a large scatter, generally pulsars with shorter periods have
larger $W_{50}$. A straight line
fit to these data has a slope of --0.29 and a width at a period of 1~s
of 4.7\degr. This is generally consistent with results already in
the literature \citep{ran93,kxl+98,mr02,wj08b}.
All other things being equal, a slope of 0 would imply that the 
emission height is a fixed fraction of the light-cylinder radius whereas a 
slope of --0.5 implies a fixed emission height in km (see e.g. the
discussion in \citealt{mr02}).
There are, however,
many additional factors which muddy the interpretiation of Figure~\ref{pw}.
These include evolution of $\alpha$ with time \citep{tm98,jk17},
the patchiness of the beam \citep{lm88}, the possibility of varying 
emission height as a function of period or age \cite{kj07}, and the variation 
of emission height as a function of longitude \citep{gg03,mr02}.

\begin{figure}
\includegraphics[width=6cm,angle=270]{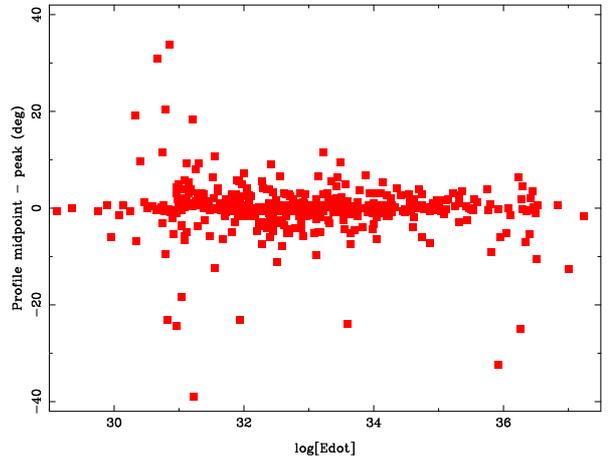}
\caption{Spin-down energy ($\dot{E}$) versus symmetry offset for 484 pulsars.}
\label{edot-sym}
\end{figure}

\begin{figure*}
\begin{center}
\begin{tabular}{cc}
\includegraphics[width=8cm,angle=0]{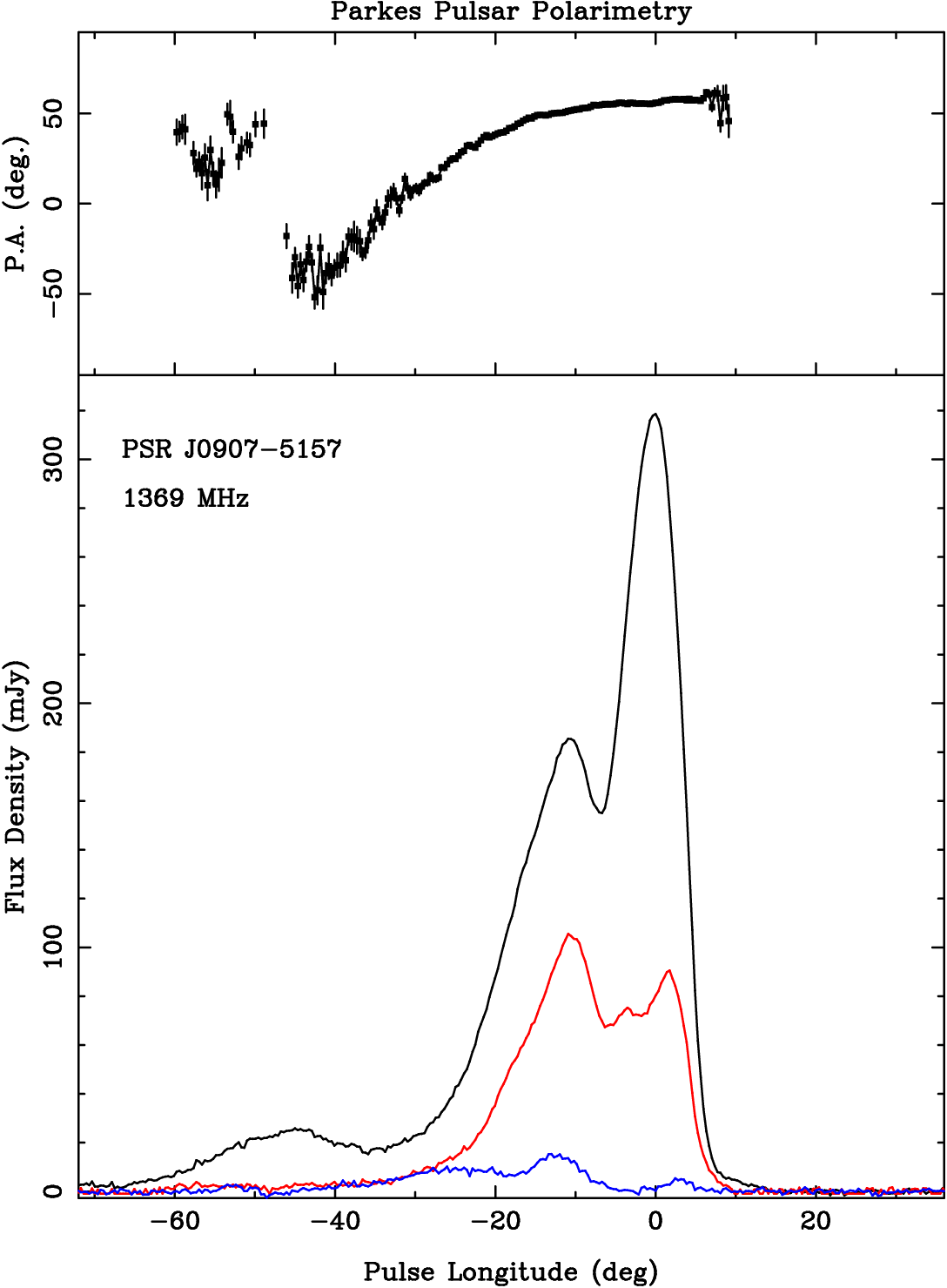} &
\includegraphics[width=8cm,angle=0]{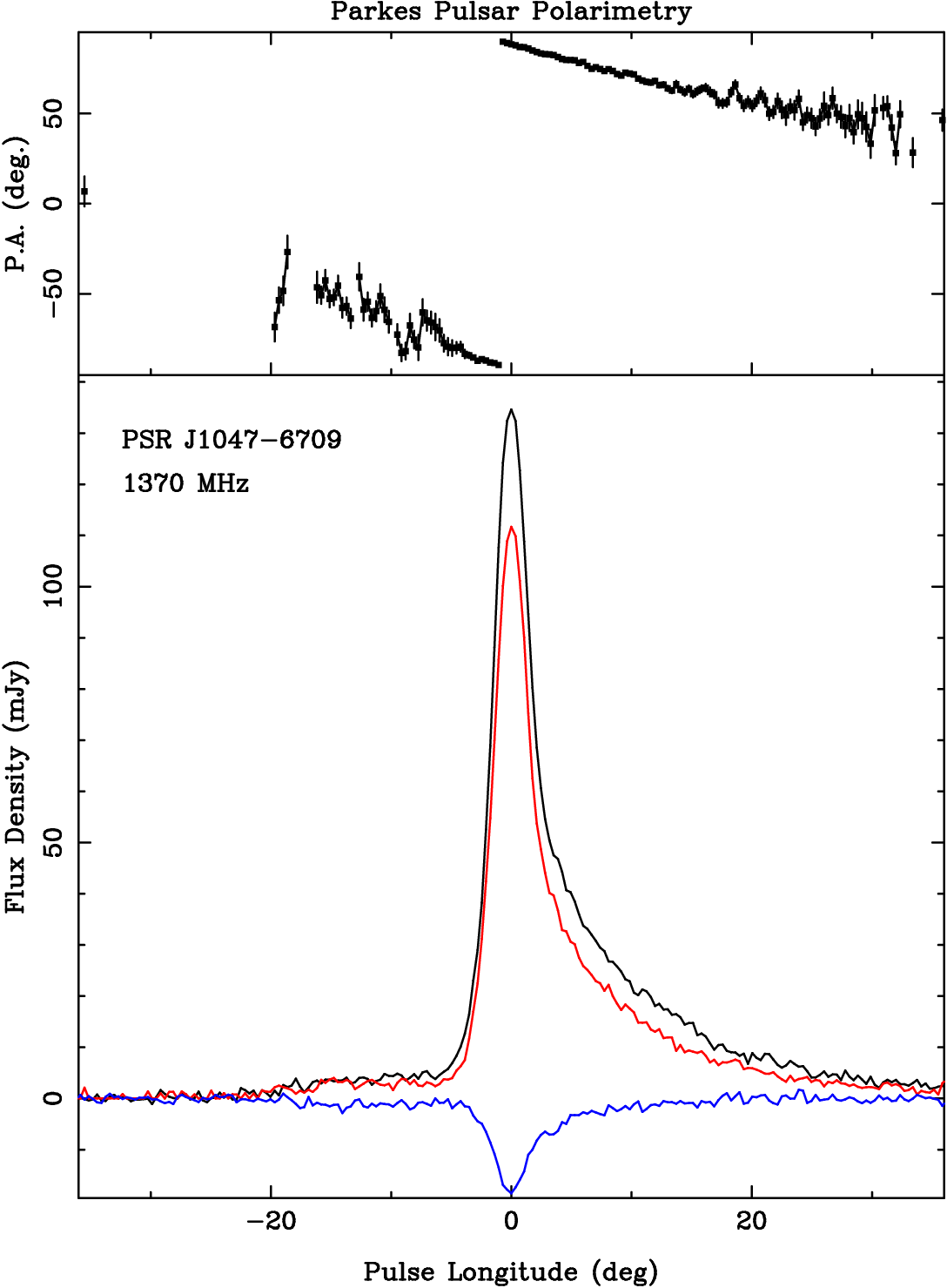} \\
\end{tabular}
\end{center}
\caption{Two examples of asymmetric profiles. Left panel is PSR~J0907--5157,
heavily weighted towards the trailing edge, right panel is PSR~J1047--6709
showing the profile dominates on the leading edge. See also the
caption to Figure~\ref{J0601}.}
\label{2psrs}
\end{figure*}

The symmetry of pulsar profiles can also be examined. In \citet{wj08b} it
was shown that there was a strong correlation between a pulsar's profile
and its time-reversed profile. Here we examine the relationship between
the peak of the total intensity profile and its midpoint. The midpoint
is defined as the half-way point between where the intensity of the leading
edge of the profile first exceeds 50\% of the maximum and where the intensity
of the trailing edge first drops below 50\% of the maximum.
Figure~\ref{edot-sym}
shows the offset between the location of the profile peak and its midpoint
as a function of $\dot{E}$ for 484 pulsars.
In almost half (237/484) of the pulsars the profile peak coincides with the
midpoint to within $\pm$~1\degr. The remaining pulsars are split evenly
between positive (peak lagging the midpoint) and negative values
although at high $\dot{E}$ a preponderance of negative values are seen
indicating the peak on the trailing part of the profile (as also pointed
out by \citealt{jw06}). This appears contrary to the cyclotron absorption
ideas of \citet{lm01} but in line with the more complete treatment
given in \citet{flm03}. Two examples of pulsars with asymmetric profiles
are given in Figure~\ref{2psrs}. Both appear to be partial cones in
the nomenclature of \citet{lm88} and there is little evolution of
the profile with observing frequency in either case. In PSR~J0907--5157
we likely observe the trailing inner and outer cones with the leading
outer cone not visible whereas for PSR~J1047--6709 only the leading
cone is detectable.


\section{Data Access}
The following data products are provided via the CSIRO's Data Access Portal.
\begin{itemize}
\item A {\sc readme} file which describes the table columns, the figure
caption and the paper reference.
\item A single {\sc psrfits} file for each of the 600 pulsars with either
512 or 1024 bins across the profile.
These files can be manipulated with the {\sc psrchive} package available
via {\sc sourceforge}. Both the {\sc psrfits} format and 
{\sc psrchive} are described in \citet{hvm04}.
The naming convention for the files is pulsarname.1400MHz.psrfits.
\item A single figure in {\sc png} format for each of the 600 pulsars similar 
to those shown in Figures~\ref{J0601} and \ref{4psrs}. The naming convention
is pulsarname.1400MHz.png.
\item A table of 600 lines (one per pulsar) with the columns as in
Table~\ref{bigtable} but in csv format with the columns separated by commas.
\end{itemize}

The data can be accessed through the main portal entry at
\url{http://data.csiro.au} and then searching on ``pulsar polarimetry" or 
the data can be accessed directly via the published DOI
\url{http://doi.org/10.4225/08/59952c840ae35}.

\section{Summary}
In this paper we provide the polarization characteristics for 600
pulsars from observations taken at 1.4~GHz using the Parkes radio
telescope since 2004. We have highlighted some of the
statistical properties of the data, including differences between
pulsars with high and low values of $\dot{E}$. All the data and
the table of parameters are made available via public archive.
A large fraction of these pulsars have also been observed at 0.7
and 3.1~GHz; future plans include a release of these data and
a full multi-frequency comparison of the population.

\begin{table*}
\caption{Pulsar polarization properties at 1.4~GHz (sample)}
\label{bigtable}
\begin{tabular}{lrrrrrrrrrrrrl}
Jname & Period & DM & RM & log($\dot{E}$) & nbin & $\sigma_I$ & $S_{1.4}$ & $W_{50}$ & $L/I$ & $V/I$ & $|V|/I$ & err & flag \\
& (ms) & (cm$^{-3}$pc) & (radm$^{-2}$) & (ergs$^{-1}$) & & (mJy) & (mJy) & (deg) & \% & \% & \% & \% \\
\hline & \vspace{-3mm} \\
J0034-0721 &    943.0 &     10.9 &      9.8 &     31.3 & 1024 &     0.27 &    11.12 &     14.4 &     10.7 &      7.7 &      7.5 &      3.0 &  \\ 
J0051+0423 &    354.7 &     13.9 &    -30.5 &     30.8 & 512 &     0.39 &     0.49 &     26.0 &     13.1 &     -2.3 &     11.2 &      3.3 &  \\ 
J0108-1431 &    807.6 &      2.4 &     -0.3 &     30.8 & 1024 &     0.57 &     1.37 &     11.6 &     76.7 &     15.5 &     13.1 &      3.1 &  \\ 
J0134-2937 &    137.0 &     21.8 &     13.0 &     33.1 & 1024 &     0.37 &     3.90 &      4.9 &     45.3 &    -17.2 &     16.9 &      3.0 &  \\ 
J0151-0635 &   1464.7 &     25.7 &     -7.0 &     30.7 & 512 &     1.01 &     1.57 &     28.8 &     29.1 &     -1.7 &      4.2 &      3.3 &  \\ 
J0152-1637 &    832.7 &     11.9 &      2.0 &     31.9 & 1024 &     0.44 &     2.14 &      6.3 &     15.1 &      1.1 &      6.0 &      3.0 &  \\ 
J0206-4028 &    630.6 &     12.9 &     -4.0 &     32.3 & 1024 &     0.51 &     0.75 &      2.8 &     10.6 &      9.3 &      9.9 &      3.1 &  \\ 
J0211-8159 &   1077.3 &     24.4 &     54.0 &     31.0 & 512 &     0.77 &     0.33 &     16.2 &     17.0 &     11.7 &     15.4 &      5.5 & W \\ 
J0255-5304 &    447.7 &     15.9 &     32.0 &     31.1 & 1024 &     0.65 &     5.00 &      6.0 &      7.3 &     -4.1 &      5.5 &      3.0 &  \\ 
J0304+1932 &   1387.6 &     15.7 &     -8.3 &     31.3 & 1024 &     0.92 &    15.05 &     10.9 &     33.4 &     15.1 &     14.8 &      3.0 &  \\ 
J0343-3000 &   2597.0 &     20.2 &     33.7 &     29.1 & 1024 &     0.96 &     1.26 &      4.9 &     14.3 &      3.1 &      3.9 &      3.2 &  \\ 
J0401-7608 &    545.3 &     21.6 &     19.0 &     32.6 & 1024 &     0.42 &     3.80 &     13.4 &     28.6 &     -0.1 &      4.7 &      3.0 &  \\ 
J0448-2749 &    450.4 &     26.2 &     24.0 &     31.8 & 1024 &     1.01 &     2.04 &      8.4 &     23.9 &    -13.3 &     11.8 &      3.0 &  \\ 
J0450-1248 &    438.0 &     37.0 &     13.0 &     31.7 & 512 &     1.01 &     1.22 &     20.4 &     25.3 &      2.5 &      6.0 &      3.4 &  \\ 
J0452-1759 &    548.9 &     39.9 &     13.8 &     33.1 & 1024 &     0.47 &    16.83 &     17.6 &     18.9 &      3.6 &      4.2 &      3.0 &  \\ 
J0459-0210 &   1133.1 &     21.0 &     18.0 &     31.6 & 1024 &     1.50 &     0.63 &      3.2 &     10.4 &    -12.9 &      9.6 &      3.8 &  \\ 
J0520-2553 &    241.6 &     33.8 &     19.0 &     31.9 & 512 &     0.96 &     0.79 &     14.1 &     18.2 &     -4.3 &      5.0 &      3.6 &  \\ 
J0525+1115 &    354.4 &     79.4 &     37.0 &     31.8 & 1024 &     0.58 &     1.74 &     14.1 &     10.6 &     12.5 &     15.5 &      3.0 &  \\ 
J0528+2200 &   3745.5 &     50.9 &    -40.2 &     31.5 & 2048 &     2.48 &     8.90 &     14.6 &     36.9 &     -4.9 &      4.6 &      3.0 &  \\ 
J0533+0402 &    963.0 &     83.7 &    -71.2 &     30.9 & 1024 &     0.86 &     0.73 &      5.6 &     13.3 &      4.0 &      5.5 &      3.2 &  \\ 
J0536-7543 &   1245.9 &     17.5 &     25.2 &     31.1 & 1024 &     0.58 &     8.42 &     16.5 &     48.8 &    -11.1 &     11.0 &      3.0 &  \\ 
J0540-7125 &   1286.0 &     29.4 &     43.0 &     31.2 & 512 &     0.94 &     0.35 &      8.4 &     14.2 &      3.1 &     15.8 &      4.8 &  \\ 
J0543+2329 &    246.0 &     77.7 &      8.7 &     34.6 & 1024 &     0.53 &    12.90 &      7.7 &     45.2 &     -8.2 &      7.9 &      3.0 &  \\ 
J0601-0527 &    396.0 &     80.5 &     64.0 &     32.9 & 1024 &     0.62 &     2.59 &     14.4 &     30.9 &      4.4 &     11.3 &      3.0 &  \\ 
J0614+2229 &    335.0 &     96.9 &     69.0 &     34.8 & 1024 &     0.73 &     3.60 &      6.7 &     72.0 &     20.3 &     20.1 &      3.0 &  \\ 

\hline
\end{tabular}
\end{table*}

\section*{Acknowledgments}
We thank G. Hobbs, A. Karastergiou, B. Koribalski and M. Kramer for useful
discussions, C. Tiburzi for her datasets and 
R. Manchester for providing RMs prior to publication.
L. Toomey provided valuable assistance with setting up the data archive.
We thank the referee for a careful reading of the manuscript.
The ATNF pulsar catalogue at 
\url{http://www.atnf.csiro.au/people/pulsar/psrcat/}
was used for this work.
The Parkes telescope is part of the Australia Telescope National Facility 
which is funded by the Commonwealth of Australia for operation as a 
National Facility managed by CSIRO. Work at NRL is supported by NASA.

\bibliographystyle{mnras}
\bibliography{poln}
\bsp
\label{lastpage}
\end{document}